\documentclass[11pt]{article}

\usepackage{epsfig}
\usepackage{amsfonts,amssymb}

\font\mybb=msbm10 at 12pt
\def\bb#1{\hbox{\mybb#1}}
\def\bZ {\bb{Z}}
\def\bR {\bb{R}}

\newcommand{\del}{\ensuremath{\partial}}

\newcommand{\half}{\ensuremath{\frac{1}{2}}}

\newcommand{\be}{\begin{equation}}
\newcommand{\ee}{\end{equation}}

\newcommand{\ba}{\begin{eqnarray}}
\newcommand{\ea}{\end{eqnarray}}
\newcommand{\gsim}{\raise.3ex\hbox{$>$\kern-.75em\lower1ex\hbox{$\sim$}}}
\newcommand{\lsim}{\raise.3ex\hbox{$<$\kern-.75em\lower1ex\hbox{$\sim$}}}

\begin{document}
\bigskip
\hskip 4.8in\vbox{\baselineskip12pt
\hbox{hep-th/0302nnn} \hbox{SUSX-TH/03-007} \hbox{LPT-ORSAY 03/50} }

\bigskip
\bigskip
\bigskip

\begin{center}
{\Large \bf Singular tachyon kinks from regular profiles}\\
\end{center}

\bigskip
\bigskip
\bigskip

\centerline{\bf E.~J.~Copeland$^\natural$, P.~M.~Saffin$^\natural$
                and D.~A.~Steer$^\sharp$}

%\bigskip
%\bigskip
\bigskip

\centerline{\it $^\natural$Centre for Theoretical Physics}
\centerline{\it University of Sussex, Falmer, Brighton BN1 9QJ,
U.K.}
\centerline{\small \tt E.J.Copeland@sussex.ac.uk}
\centerline{\small \tt P.M.Saffin@sussex.ac.uk}
\centerline{$\phantom{and}$} \centerline{\it $^\sharp$
            Laboratoire de Physique Th\'eorique
            \footnote{Unit\'e Mixte de Recherche du CNRS (UMR 8627).},
            B\^at. 210,}
\centerline{\it Universit\'e Paris XI, 91405 Orsay Cedex, France}
\centerline{and}
\centerline{\it F\'ed\'eration de recherche APC,
            Universit\'e Paris VII,}
\centerline{\it 2 place Jussieu, 75251 Paris Cedex 05, France.}
\centerline{\small \tt Daniele.Steer@th.u-psud.fr}

%\bigskip
%\bigskip

\begin{abstract}
We demonstrate how Sen's singular kink solution of the
Born-Infeld tachyon action can be constructed by taking the
appropriate limit of initially regular profiles. It is shown that
the order in which different limits are taken plays an important
r\^ole in determining whether or not such a solution is obtained
for a wide class of potentials.  Indeed, by introducing a small
parameter into the action, we are able circumvent the results of a
recent paper which derived two conditions on the asymptotic
tachyon potential such that the singular kink could be recovered
in the large amplitude limit of periodic solutions.  We show that
this is explained by the non-commuting nature of two limits, and
that Sen's solution is recovered if the order of the limits is
chosen appropriately.
\end{abstract}

%%%%%%%%%%%%%%%%%%%%%%%%%%%%%%%%%%%%%%%%%%%%%%%%%%%%%%%%%%%
\section{Introduction}
%%%%%%%%%%%%%%%%%%%%%%%%%%%%%%%%%%%%%%%%%%%%%%%%%%%%%%%%%%%
It is by now well known that the spectrum of objects in string theory
includes unstable, uncharged non-BPS D-branes \cite{sen:98a}-\cite{bergshoeff:00}.
The dynamics of these unstable D-branes is determined by an action
similar to that of D-branes, with the addition of a tachyon scalar
that indicates the presence of an instability.
Ignoring all fields other that the tachyon $T$,
the action for a single non-BPS D$p$-brane
%with $p$ spatial dimensions
is of the Born-Infeld form
\be
S=-\int {\rm d}^{p+1}x V(T)\sqrt{-{\rm det}(\eta_{\mu \nu} +
2\pi\alpha'\del_\mu T\del_\nu T)}, \label{orig-a}
\ee
where $\eta_{\mu \nu}$ is the Minkowski metric with signature
$(-,+,+,\ldots)$, and $V(T)$ the tachyon potential which has a
global maximum at $T=0$. Since a possible endstate is the vacuum,
the minimum of the potential must be at zero and this is reached
when $|T| \rightarrow \infty$. Another possible end state is a
stable D$(p-1)$ brane. This is understood \cite{sen:03} to be a
static and stable tachyon kink, and enforces a $\bZ_2$ symmetry of
the potential. The formation of singular kinks as a dynamical
process in finite time has been looked at in \cite{cline:03}.

Derrick's theorem suggests that the kink solutions of this action
are singular \cite{Sen0,Min,Lam,brax:03}.  (Note that since the
vacuua are at $|T| \rightarrow \infty$, the topological kink has
an infinite amplitude.) If such a singular solution could be shown
to exist as a limit of a regular solution, this would give us more
confidence in addressing its behaviour and interpretation. In
reference \cite{brax:03}, one of us showed how this singular
solution could be generated from a sequence of regular solutions
as long as the potential $V(T)$ satisfied two restrictions
involving its slope as $|T| \rightarrow \infty$. In reaching this
conclusion, the manifold on which the solution is defined was
altered from the line $\bR^1$ to a circle $S^1$ thereby
guaranteeing the finiteness of the energy, and hence establishing
the validity of Derricks's argument to this case. The idea was
then to take the limit where the circle became large and the
amplitude of the kink tended to infinity in order to regain the
single kink solution of Sen.  The two conditions on
$V(|T|\rightarrow \infty)$ found in \cite{brax:03} arose precisely
in order that this regularisation scheme should give the single
singular kink solution.
%
%However, it was shown that such a limiting procedure places strong
However, these two conditions on $V(T)$ contrast somewhat with
the original idea of Sen in which the single kink solution was, to
a large extent, independent of the form of the potential (having
to satisfy simple asymptotic properties only).

In the rest of this short note, we follow the spirit of
\cite{brax:03}, but show that if (as opposed to \cite{brax:03}) we
start with a slightly {\it modified} action we can recover, in a
controlled way, the singular kink solution from a series of
regular solutions (again compactified on $S^1$) without placing
restrictions on the form of the potential. The explanation for the
different result to \cite{brax:03} lies in the subtleties
associated with two different limits which do not commute.  This
situation should be contrasted with that which occurs when
studying regular solitons such as those in $\lambda \phi^4$
theory: in that case there is no problem with non-commuting
limits.

%%%%%%%%%%%%%%%%%%%%%%%%%%%%%%%%%%%%%%%%%%%%%%%%%%%%%%%%%%%
\section{The kink solution}
%%%%%%%%%%%%%%%%%%%%%%%%%%%%%%%%%%%%%%%%%%%%%%%%%%%%%%%%%%%

We look for static solutions to the tachyon equations coming
from the action (\ref{orig-a}), so that
the energy functional is
\ba
\label{EE}
E[T]&=&\int {\rm d}^{p}x V(T)\left[
        \sqrt{1+{\del}_i T{\del}^i T}\right] \; , \qquad (2\pi\alpha'=1).
\ea
Following Derrick's theorem \cite{Raj},
now assume that there exists a kink profile which extremizes the energy and
consider the configuration obtained by dilating the
coordinates, $x\rightarrow \lambda x$. If a solution is
to be an extremum of the energy then
\mbox{$(\del E_\lambda/\del\lambda)_{\lambda=1}=0$}, so that
\ba
\int {\rm d}^{p}x V(T)\left[
   (1+\del_i T\del^i T)^{-1/2}
   \left[p+(p-1)\del_i T\del^i T\right] \right]=0.
\ea
For $p= 1$ this forces $V=0$ or $\del_i T\del^i T\rightarrow\infty$.
This is the reasoning
behind the singular kink solution \cite{Sen0}
\ba
\label{SenKink}
T(x) = \left\{
\begin{array}{cl}
\infty & x > 0 \\
0 & x=0 \\
-\infty & x < 0
\end{array}
\right.
\ea
whose energy is given by
\be
\label{SenE}
E_0 = \int_{-\infty}^{\infty} {\rm d}T V(T).
\ee
($E_0$ can be obtained by formally writing the solution
(\ref{SenKink}) as $T = \lim_{C \rightarrow 0} x/C$ and
substituting into (\ref{EE}).)

The profile (\ref{SenKink}) is clearly singular, complicating the
analysis of its properties; it would therefore be useful to
describe this solution as a limit of a regular solution. One
approach, adopted in \cite{brax:03}, was to start with the given
action (\ref{orig-a}), compactify the coordinate $x$ leading to
finite energy periodic solutions for $T(x)$, and then study the
behaviour of these solutions in the decompactified limit. A
crucial parameter in that analysis was the constant integral of
motion $V_0$, since the topological kink is obtained in the limit
$V_0 \rightarrow 0$.
%(We recall that $V_0 \equiv T'(\partial L/\partial T') - L$ where $L$
%is the Lagrangian,
(The amplitude of the kink solution depends inversely on $V_0$.)
This approach led to two restrictions on the allowed form of
$V(T)$ such that the singular kink (\ref{SenKink}) with energy
(\ref{SenE}) was obtained in the $V_0 \rightarrow 0$ limit.  These
conditions were that for $|T| \rightarrow \infty$, $V'/V
\rightarrow 0$ and $V'/V^2 \rightarrow \infty$.

Here we propose a different way to regularize the singular
solution. It is based on adding a small correction to the action
(\ref{orig-a}). With that correction, {\it for all} $V(T)$ we can
recover the single kink solution (\ref{SenKink}) with energy
(\ref{SenE}) (provided of course that the integral (\ref{SenE}) is
finite). Then, at the end, we take the limit where the correction
vanishes. As can be anticipated when trying to construct such
singular kinks, there is an issue of non-commuting limits (see the
discussion in the next section).

Before proceeding, note that in the case of regular kinks, such
as the tanh-like solutions of $\lambda \phi^4$ theory in 1+1
dimensions, both the two methods outlined in the previous
paragraph can be used to construct the topological kink.  They
both give the correct result independently of the order in which
limits are taken.

Consider therefore the modified action and energy functional
\ba
S & = & \int {\rm d}^{p+1}x {\cal L} = -\int {\rm d}^{p+1}x
V(T)\left[ \sqrt{1 + \del_\mu T\del^\mu T} + \epsilon (\del_\mu
T\del^\mu T)^n \right] \label{modA},
\\
E[T]&=&\int {\rm d}^{p}x V(T)\left[
        \sqrt{1+\del_i T\del^i T}
        +\epsilon \left(\del_i T\del^i T\right)^n \right]
\label{modaction},
\ea
where $n>\half$. Note that taking the small parameter
$\epsilon\rightarrow 0$ gives back the original action
(\ref{orig-a}). We now follow the prescription in \cite{brax:03},
and repeat the scaling argument to find
\ba
\int {\rm d}^{p}x V(T)\left[
   (1+\del_i T\del^i T)^{-1/2}
   \left[p+(p-1)\del_i T\del^i T\right]
   +(p-2n)\epsilon \left(\del_i T\del^i T\right)^n\right]=0.
\label{con}
\ea
For $p=1$ this again leads to the singular kink (\ref{SenKink})
when $\epsilon \rightarrow 0$. For $\epsilon \neq 0$, however,
equation (\ref{con}) can now be satisfied by
\ba
T'^{4n+2}+T'^{4n}=\frac{1}{(2n-1)^2\epsilon^2},
\ea
(the dash denotes a derivative with respect $x$), whose solution
is a linear tachyon profile $T(x)=\kappa x$. The singular kink is
recovered in the limit $\epsilon\rightarrow 0$, in which case
$\kappa^{2n+1}\rightarrow 1/[(2n-1)\epsilon] \rightarrow \infty$.
Using the same arguments as those which led to (\ref{SenE}), the
energy of this solution is given by
\ba
E_{\epsilon\rightarrow 0}=E_{0}
     \left\{1+1/[(2n-1)\kappa^2]\right\}
\ea
for all values of $n>\half$. This reproduces the energy of a single kink
as
\mbox{$1/\kappa^2\rightarrow[(2n-1)\epsilon]^{2/(2n+1)}\rightarrow
0$}. While Derrick's theorem shows that a solution could exist, it
is not guaranteed. However, the equations of motion coming from
(\ref{modA}) are (denoting $V'=\del V/\del T$),
%\ba
%\left[(1+\del_\mu T\del^\mu T)^{-3/2}+2n(2n-1)\epsilon (\del_\mu
%T\del^\mu T)^{(n-1)} \right]\del_\mu\del^\mu T
%  &=&\frac{V'}{V}\left[ (1+\del_\mu T\del^\mu T)^{-1/2}
%\right. \nonumber \\
%                      &~& \left. -(2n-1)\epsilon (\del_\mu T\del^\mu T)^{n}\right]
%\ea
\ba
\left[(1+T'^2)^{-3/2}+2n(2n-1)\epsilon T'^{2(n-1)} \right]T''
  &=&\frac{V'}{V}\left[ (1+T'^2)^{-1/2}-(2n-1)\epsilon T'^{2n}\right]
\ea
showing that for $p=1$ the static, linear profile is indeed a
solution.
Furthermore by doing a perturbative analysis, one can
show that it is a stable solution.  For completeness we note that
for the static kink $V_0$ is given by
\ba
V_0 \equiv T'\frac{\partial {\cal L}}{\partial T'} - {\cal L} =
V(T) \left[ (1+T'^2)^{-1/2} -(2n-1)\epsilon T'^{2n}\right].
\ea
As we shall be studying oscillatory solutions note that at $T'=0$
(the peaks of the wave) we have $V_0=V(T_{\rm peak})$, and as $V(T)$ decrease
for large $|T|$ then large amplitude solutions have small $V_0$.

A slightly different way of achieving similar results is the following.
Rather than taking the action modified
according to (\ref{modA}), consider instead
\be
S = - \int  {\rm d}^{p+1}x V(T) ( 1 + \del_\mu T\del^\mu T)^{q}.
\label{mod2}
\ee
For $q>1/2$, the equations of motion (or equivalently a Derrick-type
argument) show that there is a topological kink solution given by
$T = x/{\sqrt{2q-1}}$.
Now follow the construction of \cite{brax:03}.  Taking $q=1/2$ first
followed by the $V_0 \rightarrow 0$ limit leads to the two conditions on
$V(T)$ discussed in \cite{brax:03}.  On the other hand, notice that the {\it
opposite} choice of limits, namely $V_0 \rightarrow 0$ followed by
$q \rightarrow 1/2^+$ gives the singular kink (\ref{SenKink}) with energy $E_0$
for all $V(T)$.  Hence we conclude that the limits
\be
q \rightarrow 1/2^+ \qquad {\rm and} \qquad  V_0 \rightarrow 0
\ee
do not commute.

%%%%%%%%%%%%%%%%%%%%%%%%%%%%%%%%%%%%%%%%%%%%%%%%%%%%%%%%%%%
\section{A question of limits}
%%%%%%%%%%%%%%%%%%%%%%%%%%%%%%%%%%%%%%%%%%%%%%%%%%%%%%%%%%%
We have seen that the single kink can be found from the action
Eqn.~(\ref{modaction}) once the appropriate limit for $\epsilon$
is taken. Similarly it can be obtained from the action
(\ref{mod2}) again being careful with the order of limits.
Indeed, as opposed to regular solitons, it seems crucial to start
with a {\it modified} action in order to be able to reproduce the
singular kink for all $V(T)$. To show explicitly this
non-commuting nature, we look for regular kink solutions on a
circle using the regularized energy functional
Eqn.~(\ref{modaction}). As opposed to \cite{brax:03} we take the
decompactified and $V_0 \rightarrow 0$ limits {\it followed} by
$\epsilon \rightarrow 0$. The typical solutions are periodic, with
the wavelength representing the separation of the kinks and
anti-kinks. To reproduce the single kink (\ref{SenKink}) we need a
solution with infinite wavelength i.e. infinite kink/anti-kink
separation, and infinite amplitude as discussed above.

To be specific, we take the exponential potential $V=\exp(-T^2)$
as an example, but the results are independent of this specific
form. Note that this potential does {\it not} satisfy the conditions of
\cite{brax:03}. Typical periodic solutions for $\epsilon=0$ are
shown in Fig.\ 1. Here we see the problem which was examined in
\cite{brax:03}, namely that as the amplitude increases (i.e.\
$V_0$ decreases), the wavelength decreases. This behaviour means
that with $\epsilon=0$ one cannot reproduce (\ref{SenKink}) in the
large radius (large wavelength) limit. In contrast to this, Fig.\
2 shows some typical profiles for $\epsilon \neq 0$. What we see
here is that the large amplitude solutions track the regular
single kink solution as $T$ increases from zero.  In other words
increasing amplitude solutions have increasing wavelength --- the
type of behaviour required to reproduce the result of
(\ref{SenKink}) for all $V(T)$. As the regular single kink
solution exists for all non-zero $\epsilon$ this behaviour
persists as $\epsilon$ is reduced.

From the perspective of the regularized action we now see that the
constraints on $V(T)$ discussed in \cite{brax:03} arise when the
$\epsilon \rightarrow 0$ limit is taken {\it before} the large
amplitude (or $V_0 \rightarrow 0$) limit.  Here we take the $V_0
\rightarrow 0$ limit followed by $\epsilon \rightarrow 0$ limit.
In fact we can go further and  consider the relationship between
all three quantities, $\epsilon$, wavelength and amplitude. Fig.\
3 allows us to follow the contours of constant wavelength as we
reduce $\epsilon$. They naturally lead to regions of increasing
amplitude, and in the limit $\epsilon\rightarrow 0$ following such
a contour we arrive at an array of singular kinks/anti-kinks with
a separation depending on the contour chosen.

%%%%%%%%%%%%%%%%%%%%%%%%%%%%%%%%
\begin{figure}
\center \epsfig{file=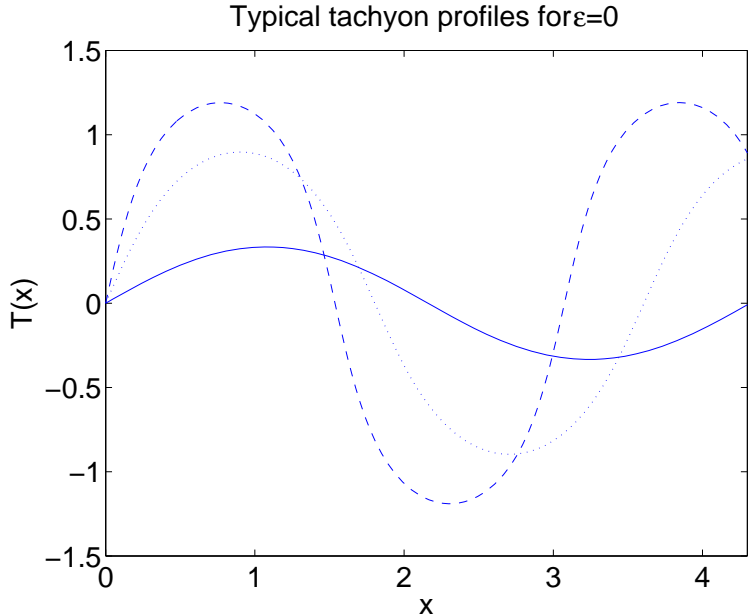,width=10cm} \flushleft
\caption{Periodic tachyon profiles $T(x)$ for the potential
$V=e^{-T^2}$ and $\epsilon=0$.  The curves have increasing
amplitude or equivalently decreasing $V_0$.}
\end{figure}
%%%%%%%%%%%%%%%%%%%%%%%%%%%%%%%%
%%%%%%%%%%%%%%%%%%%%%%%%%%%%%%%%
\begin{figure}
\center \epsfig{file=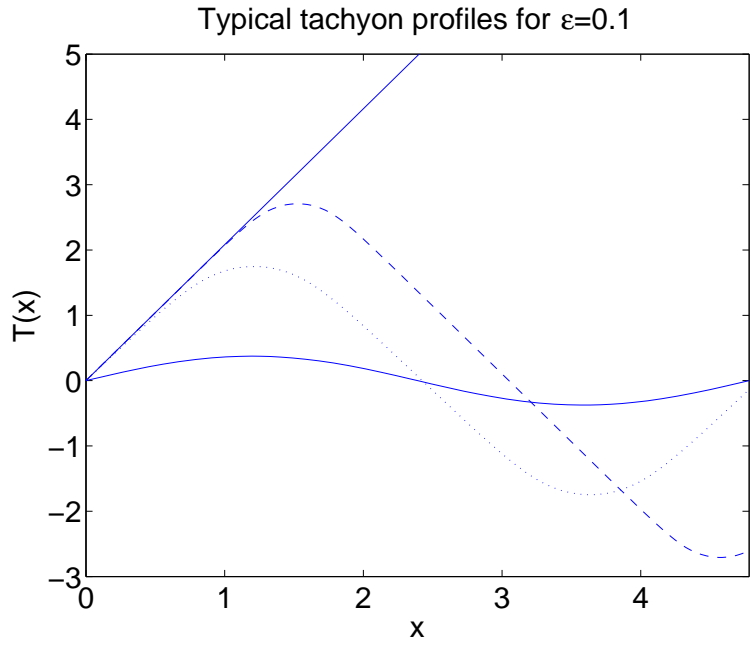,width=10cm} \flushleft
\caption{Periodic tachyon profiles $T(x)$ for the potential
$V=e^{-T^2}$ and $\epsilon=0.1$.  The curves have increasing
amplitude. The straight line corresponds to the regular single
kink solution given in the text.}
\end{figure}
%%%%%%%%%%%%%%%%%%%%%%%%%%%%%%%%
%%%%%%%%%%%%%%%%%%%%%%%%%%%%%%%%
\begin{figure}
\center
\epsfig{file=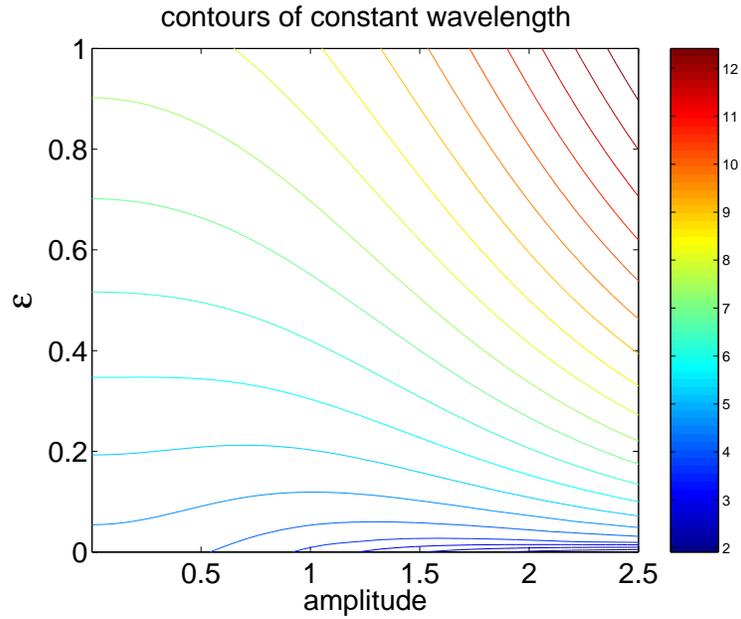,width=10cm}
\flushleft
\caption{The dependence of wavelength on $\epsilon$ and amplitude.}
\end{figure}
%%%%%%%%%%%%%%%%%%%%%%%%%%%%%%%%

%%%%%%%%%%%%%%%%%%%%%%%%%%%%%%%%%%%%%%%%%%%%%%%%%%%%%%%%%%%
\section{Conclusions}
%%%%%%%%%%%%%%%%%%%%%%%%%%%%%%%%%%%%%%%%%%%%%%%%%%%%%%%%%%%
In this note, motivated by the results of \cite{brax:03}, we have
attempted to address the question of whether it is possible to
produce the singular kink solution of Sen by taking the
appropriate limit of a regular class of solutions. The original
work of Sen suggests that the solution exists subject to only mild
constraints on the tachyon potential $V(T)$. We have shown how it
is possible to reproduce the single kink solution, provided we
added a term to the action and later send it to zero.  This allows
the singular kink to be reached in a controlled manner from
regular solutions, without the restrictions found in
\cite{brax:03}. This result is important because it lifts the
rather tight restrictions on the tachyon potential found in
\cite{brax:03} and brings the requirements on $V(T)$ more in line
with those originally anticipated by Sen. However, it also
underlines the fact that one must treat the action (\ref{orig-a})
with some care.

%%%%%%%%%%%%%%%%%%%%%%%%%%%%%%%%%%%%%%%%%%%%%%%%%%%%%%%%%%%
\section*{Acknowledgements}
We thank N.Antunes, Ph.Brax and J.Mourad for useful discussions.
DAS thanks the Particle Physics and Astronomy groups at Sussex
for support and wonderful hospitality. PMS is supported by PPARC.
%%%%%%%%%%%%%%%%%%%%%%%%%%%%%%%%%%%%%%%%%%%%%%%%%%%%%%%%%%%

%%%%%%%%%%%%%%%%%%%%%%%%%%%%%%%%%%%%%%%%%%%%%%%%%%%%%%%%%%%
%%%%%%%%%%%%%%%%%%%%%%%%%%%%%%%%%%%%%%%%%%%%%%%%%%%%%%%%%%%


\begin{thebibliography}{10}

\bibitem{sen:98a}
A. Sen.
\newblock ``Tachyon Condensation on the Brane Antibrane System'',
\newblock {\em JHEP.} 08,012 (1998),
\newblock hep-th/9805170.

\bibitem{sen:98b}
A. Sen.
\newblock ``SO(32) Spinors of Type I and Other Solitons on Brane-Antibrane
Pair'',
\newblock {\em JHEP.} 09,023 (1998),
\newblock hep-th/9808141.

\bibitem{horava:98}
P. Horava.
\newblock ``Type IIA D-Branes, K-Theory, and Matrix Theory'',
\newblock {\em Adv.Theor.Math.Phys.} 2,1373 (1999),
\newblock hep-th/9812135.

\bibitem{sen:99}
A. Sen.
\newblock ``Non-BPS states and branes in string theory'',
\newblock hep-th/9904207.

\bibitem{sen:99b}
A. Sen.
\newblock ``Supersymmetric world volume action for non-BPS D-branes.'',
\newblock {\em JHEP.} 9910:008 (1999)
\newblock hep-th/9909062

\bibitem{sen:03}
A. Sen.
\newblock ``Open and Closed Strings from Unstable D-branes'',
\newblock hep-th/0305011.

\bibitem{garousi:00}
M. Garousi.
\newblock ``Tachyon couplings on non-BPS D-branes and Dirac-Born-Infeld action'',
\newblock {\em Nucl. Phys. B584}, 284 (2000).
\newblock hep-th/003122.

\bibitem{garousi:02}
M. Garousi.
\newblock ``T-duality and Actions for Non-BPS D-branes'',
\newblock {\em Nucl. Phys. B647}, 117 (2002).
\newblock hep-th/0209068.

\bibitem{bergshoeff:00}
E. Bergshoeff, M. de Roo, T. de Wit, E. Eyras, S. Panda.
\newblock ``On-shell S-matrix and tachyonic effective actions'',
\newblock {\em JHEP.} 0005:009 (2000) 
\newblock hep-th/0003221.

\bibitem{cline:03}
J. Cline, H. Firouzjahi.
\newblock ``Real-time D-brane condensation.'',
\newblock hep-th/0301101.

%\cite{Sen:2003tm}
\bibitem{Sen0}
A.~Sen,
``Dirac-Born-Infeld action on the tachyon kink and vortex,''
arXiv:hep-th/0303057.
%%CITATION = HEP-TH 0303057;%%

\bibitem{Min}
J.~A.~Minahan and B.~Zwiebach,
%``Gauge fields and fermions in tachyon effective field theories,''
JHEP {\bf 0102}, 034 (2001), arXiv:hep-th/0011226.
%%CITATION = HEP-TH 0011226;%%

\bibitem{Lam}
N.~D.~Lambert and I.~Sachs,
%``Tachyon dynamics and the effective action approximation,''
{\it Phys.\ Rev.\ D} {\bf 67}, 026005 (2003),
arXiv:hep-th/0208217.
%%CITATION = HEP-TH 0208217;%%

\bibitem{brax:03}
P. Brax, J. Mourad and D. Steer.
\newblock ``Tachyon kinks on non-BPS D-branes'',
\newblock hep-th/0304197 .

\bibitem{Raj}
R.~Rajaraman, ``Solitons And Instantons. An Introduction To
Solitons And Instantons In Quantum Field Theory,'' {\it
Amsterdam, Netherlands: North-holland (1982) 409p}.



\end{thebibliography}
\end{document}